\documentclass[a4,12pt]{article}
\usepackage{amsmath}
\usepackage{graphicx}
\usepackage{amssymb} 
\newcommand{\Mat}[1]{{{\boldsymbol{#1}}}}
\newcommand{\abs}[1]{\left\vert#1\right\vert}
\def\be{\begin{equation}}
\def\ee{\end{equation}}
\def\dd{\mathrm{d}}
\date{}
\begin{document}
\title{\bf Space isotropy and weak equivalence principle in a scalar theory of gravity}
\author{M. Arminjon}
%
%

\maketitle
\begin{center} 
{ \small Laboratoire ``Sols, Solides, Structures'' \\
(CNRS, Universit\'e Joseph Fourier, \& Institut National Polytechnique de Grenoble) 
\\ BP 53, F-38041 Grenoble cedex 9, France. \\
Currently at Dipartimento di Fisica/ Universit\`a di Bari, \& INFN/ Sezione di Bari,\\
Via Amendola 173, I-70126 Bari, Italy.}



\begin{abstract}
We consider a preferred-frame bimetric theory in which the scalar gravitational field both influences the metric and has direct dynamical effects. A modified version (``v2") is built, by assuming now a locally-isotropic dilation of physically measured distances, as compared with distances evaluated with the Euclidean space metric. The dynamical equations stay unchanged: they are based on a consistent formulation of Newton's second law in a curved space-time. To obtain a local conservation equation for energy with the new metric, the equation for the scalar field is modified: now its l.h.s. is the flat wave operator. Fluid dynamics is formulated and the asymptotic scheme of post-Newtonian approximation is adapted to v2. The latter also explains the gravitational effects on light rays, as did the former version (v1). The violation of the weak equivalence principle found for gravitationally-active bodies at the point-particle limit, which discarded v1, is proved to not exist in v2. Thus that violation was indeed due to the anisotropy of the space metric assumed in v1.
\end{abstract}
\end{center}

\section{Introduction}

The universality of gravitation, {\it i.e.}, the fact that ``all bodies fall in the same way in a gravitational field," is a distinctive feature of the gravity interaction. It is also known as the ``weak equivalence principle" (WEP), the equivalence being between the gravitational and inertial effects: indeed, the fact that the gravitational acceleration of a small body is the same independently of its composition and its mass, allows to incorporate in it the acceleration due to the motion of the arbitrary reference frame through an inertial frame. Following Galileo, Newton, and E\"otv\"os, the empirical validity of the WEP has been checked with increasing precisions \cite{Fischbach}. Therefore, the WEP is involved in the very construction of relativistic theories of gravitation, in particular in the construction of Einstein's general relativity (GR), which takes benefit of the universality of gravity to geometrize it. More precisely, in the construction of a relativistic theory, one ensures the validity of the WEP for {\it test particles}, which, by definition, do not influence the gravitational field. The geometrization operated by GR is one way to do this. Another way is to formulate the dynamics by an extension of Newton's second law, the gravitational force being $m\mathbf{g}$ with $m$ the (velocity-dependent) inertial mass and $\mathbf{g}$ a (theory-dependent) gravity acceleration \cite{A16}. We would like to emphasize that Einstein's geometrized dynamics, in which test particles follow geodesic lines of the space-time metric, {\it can be rewritten in that way} \cite{A16}. \\

A theory of gravitation has been built, in which one starts from a simple form for the field $\mathbf{g}$, suggested by a heuristic interpretation of gravity as Archimedes' thrust due to the space variation of the ``ether pressure" $p_e$, and in which $p_e$ also determines the space-time metric. One obtains a scalar bimetric theory with a preferred reference frame. This theory is summarized in Ref. \cite{B21}, or in more detail in Ref. \cite{A34}. It admits a ``physically observable preferred foliation" \cite{PittsSchieve}, in short a detectable ether. Yet, apart from the WEP violation to be discussed below, it seems to agree with observations \cite{B21,A34}. In particular, the ether of that theory could be indeed detected by adjusting on astrodynamical observations the equations of celestial mechanics that are valid in the theory \cite{B21}. One of the motivations to investigate a preferred-frame theory is that the existence of a preferred reference frame, involving that of a preferred time, may be regarded as a possible way to make quantum theory and gravitation theory match together \cite{B21}. It is interesting to note that, currently, a detectable ether is advocated already in the absence of gravitation by Selleri, in the framework of a theory close to special relativity, but based on an absolute simultaneity \cite{SelleriFP96}. \\

As we mentioned, the WEP is valid for any test particle, by construction---in the investigated scalar theory just as well as in GR. But any real body, however small it may be, must influence the gravitational field, so that one can {\it a priori} expect the occurrence of self-accelerations. The latter ones are excluded from Newton's theory by the actio-reactio principle, but this principle cannot even be formulated in a nonlinear theory---as are most relativistic theories. Moreover, the mass-energy equivalence, which has to be accounted for by a relativistic theory, implies that the rest-mass, kinetic, and gravitational energies of the test body all influence the gravitational field. This means that the internal structure of the test body may {\it a priori} be expected to influence its motion, which would be a violation of the WEP. {\it Hence, the validity of the WEP for test particles is far from ensuring that this principle applies to real bodies}. In GR, studies of the self-force acting on an extended body can be found, e.g. \cite{MinoSasaki97,QuinnWald97,Rohrlich99,Mino05}, but they have been based on simplifying assumptions such as that of a black-hole body \cite{MinoSasaki97,QuinnWald97,Mino05}, or on linearized GR \cite{Rohrlich99}. Such assumptions do not seem very appropriate to check whether a violation of the WEP might be predicted by GR in the real world, say for an asteroid or a spacecraft in the solar system. It is well-known, since the work of Nordtvedt \cite {Nordtvedt68,Nordtvedt69}, that the WEP may be violated for extended bodies of a finite mass in some relativistic theories of gravity (see also Will \cite{Will71a,Will93}). However, the kind of violation of the WEP which we have in mind is a more severe one, that occurs even in the limit in which the size of the extended body shrinks to zero \cite{A33}. \\

In the above-mentioned scalar theory \cite{A34}, it has been proposed a rigorous framework for the study of weakly-gravitating systems \cite{A23}, in conformity with the asymptotic approximation schemes which are currently used in applied mathematics, and this scheme has been developed up to the equations of motion of the mass centers of perfect-fluid bodies \cite{A25-26,A32}. This ``asymptotic" scheme of post-Newtonian approximation (PNA) is based on a one-parameter family of initial conditions, defining a family of gravitating systems. 
\footnote{\,
A scheme similar to our ``asymptotic" scheme has been previously proposed for GR by Futamase \& Schutz \cite{FutaSchutz}, though it is based on a particular initial condition as to the space metric and its time derivative, which enforces the spatial isotropy of the metric. In any case, the work \cite{FutaSchutz} is restricted to the local equations, and this in a form which is not very explicit.
}
It is hence different from the ``standard" PNA proposed for GR by Fock \cite{Fock59} and Chandrasekhar \cite{Chandra65}, which is at the basis of a significant part of the subsequent work on relativistic celestial mechanics in GR, and in which no such family is considered. The asymptotic PNA predicts that the internal structure of the bodies definitely influences their motion in a weakly-gravitating system such as our solar system, at least for the scalar theory \cite{A32}. Moreover, by considering a family of PN systems which are identical but for the size of one of the bodies, which is a small parameter $\xi$, it has been possible to make a rigorous study of the point-particle limit. It has thus been found \cite{A33} that the internal structure of a body does influence its post-Newtonian acceleration even at the point-particle limit. It has also been investigated the particular case where, apart from the small body, there is just one massive body, which is static and spherically symmetric (SSS). In that case, the PN equation of motion of the mass center of the small body is identical to the PN equation of motion of a test particle in the corresponding SSS field, {\it apart from one structure-dependent term}. These results show a patent violation of the WEP, and one whose magnitude is likely to discard the initial version of the theory \cite{A33}. The specific reason which makes the WEP violation actually occur in the point-particle limit of the PN equations of motion has been identified: it is the anisotropy of the space metric (in the preferred reference frame of the theory). More precisely, it was assumed that there is a gravitational contraction of physical objects, only in the direction of the gravity acceleration ${\bf g}$ (see Ref. \cite{O3} and references therein), thus making the PN spatial metric depend on the derivatives $U_{,i}$ of the Newtonian potential $U$. The structure-dependence (hence the non-uniqueness) of the point-particle limit comes then \cite{A33} from the fact that the self part of the second derivatives $U_{,i,j}$ is order zero in $\xi$. 
\footnote{\
The PN spatial metric valid for the standard form of Schwarzschild's metric of GR also depends on the derivatives $U_{,i}$ (with specifically $U=GM/r$ in the case of Schwarzschild's metric). Hence, the dependence of the spatial metric on the $U_{,i}$'s should be found for general situations with ``Schwarzschild-like" gauges, i.e., with gauge conditions under which the standard form of Schwarzschild's metric is the unique SSS solution of the Einstein equations with Newtonian behaviour at spatial infinity. (It is not difficult to exhibit such gauges.) One may then ask \cite{A33} whether a similar violation of the WEP might appear also in GR, in such Schwarzschild-like gauges. This would be more difficult to check, due to the greater complexity of GR as compared with the present scalar theory, in particular due to the necessity of satisfying constraint equations.
} 
\\

Therefore, it has been began \cite{O3} to investigate the case where one assumes a locally {\it isotropic} gravitational contraction, which case is nearly as natural as the anisotropic case according to the heuristic concept that led to the scalar theory, {\it i.e.}, gravity seen as due to the heterogeneity of the field of ``ether pressure" $p_e$. It has been shown that there is some freedom left for the equation governing the field $p_e$. {\it The aim of the present paper} is to propose a definite equation for the scalar gravitational field, leading to an exact local conservation law for the total (material plus gravitational) energy; to investigate the main features of this new version of the theory; and to show that it does eliminate the WEP violation which has been found with the former version. The next Section summarizes the basic concepts of the theory, which remain true for the new version to be built in this paper. Section \ref{field+energy} precises the equation for the scalar field, in connection with the necessity of obtaining an exact energy conservation. The equations of motion of a perfect fluid are obtained in Sect. \ref{fluid}. Section \ref{PNA} is devoted to the post-Newtonian approximation and shows, in particular, that this theory predicts the same effects on light rays as the standard effects known in GR. The PN equations of motion of the mass centers are derived in Sect. \ref{EMMC}. The point-particle limit is taken in Sect. \ref{pointlimit}, and our conclusion makes Sect. \ref{conclusion}.

\section{Basic concepts} \label{basic} 

The idea according to which gravity would be Archimedes' thrust due to the macroscopic part of the pressure gradient in a fluid ``ether," and the necessity to account for special relativity, lead to set a few basic assumptions \cite{O3}, which we state directly here. 

\subsection{The preferred reference frame and the metric} \label{Metric}
The space-time manifold V is assumed to be the product $\mathsf{R} \times \mathrm{M}$, where $\mathrm{M}$ is the preferred reference body, which plays the role of Newton's absolute space. The equations of the theory are primarily written in the preferred reference frame E associated with that body. 
\footnote{
A reference frame is for us essentially a (three-dimensional) reference body, plus a notion of time. Let us begin with the viewpoint of ``space plus time," which is admissible once we take the space-time to be a product, $\mathrm{V}=\mathsf{R} \times \mathrm{M}$ ---the ``time" $T$ of an event $X =(T,{\bf x}) \in \mathrm{V}$ being thus the canonical projection of $X$ into $\mathsf{R}$. From this viewpoint, a general reference frame can be defined as a time-dependent diffeomorphism $\psi_T$ of the space M onto itself. Consider the trajectories $T\mapsto \psi_T(\mathbf{x})$, each for a fixed $\mathbf{x} \in \mathrm{M}$. As $\mathbf{x}$ varies in $\mathrm{M}$, the set of these trajectories defines a deformable body N, which is uniquely associated with the reference frame $(\psi_T)_{T\in\mathsf{R}} $. Thus, $\mathbf{x'}=\psi_T(\mathbf{x})$ describes the motion of N relative to M. For a general $\psi _T$, this body N will indeed be deformed. The most obvious reference frame is yet that one which is associated with M itself, thus: $\forall \,T, \ \psi_T = \mathrm{Id}_{\mathrm{M}}$. This is our preferred frame denoted by E. From the viewpoint of ``space-time," which is more general and which is hence also admissible in an ether theory, a reference frame is defined by a three-dimensional congruence of world lines, which defines a ``body" (as we say) or ``reference fluid" \cite{Cattaneo}. Thus the trajectories of the ``space plus time" description are replaced by world lines: $s\mapsto X(s)$. Among systems of space-time coordinates: $X \mapsto \chi^\mu(X)=y^\mu \ (\mu =0,...,3)$, coordinate systems {\it adapted} to a given reference frame are such that each world line $X(s)$ belonging to the corresponding body has constant space coordinates: $\forall s, \ \chi^i(X(s)) = y^i= \mathrm{Constant}$ $(i = 1,2,3)$. 
} \label{referenceframe}
Specifically, it is assumed that the preferred reference body M is endowed with a time-invariant Euclidean metric $\Mat{g}^0$, with respect to which M is thus a rigid body, which is assumed to fill the Euclidean space. On the ``time" component of the space-time, we have the one-dimensional Euclidean metric. Combining these metrics on the component spaces, we get a Lorentzian metric: the square scalar product of an arbitrary 4-vector $\mathsf{U}=(U^0,{\bf u})$, with ${\bf u}$ an arbitrary spatial vector ({\it i.e.}, formally, an element of the tangent space $\mathrm{TM}_{{\bf x}}$ to M at some ${\bf x}\in \mathrm{M}$), is\begin{equation} \label {flatspacetimemetric}
\Mat{\gamma}^0(\mathsf{U},\mathsf{U})=(U^0)^2-\Mat{g}^0({\bf u},{\bf u}).
\end{equation}
This is the ``background metric," which should determine the proper time along a trajectory, if it were not for the metrical effects of gravity. For this to be true, it is necessary that the canonical projection of an event $X \in\mathrm{V}$ gives in fact $x^0\equiv cT$, rather than $T$, where $T$ is indeed a time (called the ``absolute time") and $c$ is a constant---the velocity of light, as measured with ``physical" space and time standards. If on M we take Cartesian coordinates $(x^i)$ for $\Mat{g}^0$, then the space-time coordinates $(x^\mu) = (x^0,(x^i))$ are Galilean coordinates for $\Mat{\gamma}^0$ [{\it i.e.}, in that coordinates, $(\gamma^0_{\mu \nu }) = \mathrm{diag}(1, -1, -1, -1)$], which are adapted to the preferred frame E. The gravitational field is a scalar field $p_e$, the ``ether pressure," which determines the field of ``ether density" $\rho_e$ from the barotropic relationship $\rho_e=\rho_e(p_e)$. Gravity has metrical effects which occur through the ratio
\be \label{beta}
\beta(T,\mathbf{x}) \equiv \rho_e(T,\mathbf{x}) /\rho_e^\infty(T) \leq 1,
\ee
with
\be \label{rho_e_infini}
\rho_e^\infty(T) \equiv \mathrm{Sup}_{\mathbf{x}\in \mathrm{M}}\rho_e(T,\mathbf{x}). 
\ee
More precisely, the ``physical" space-time metric $\Mat{\gamma}$, that one which more directly expresses space and time measurements, is related to the background metric $\Mat{\gamma}^0$ by a dilation of time standards and a contraction of physical objects, both in the ratio $\beta$. Thus, in any coordinates $(y^\mu)$ which are adapted to the frame E and such that $y^0=x^0\equiv cT$, the line element of $\Mat{\gamma}$ writes
\begin{equation} \label {spacetimemetric}
\dd s^2 = \gamma_{\mu\nu} \dd y^\mu \dd y^\nu = \beta^2 (\dd y^0)^2 - g_{ij} \dd y^i \dd y^j,
\end{equation}
where $\Mat{g}$ is the physical space metric in the frame E. This equation implies that the scalar $\beta$ can be more operationally defined as
\be \label{beta_operatoire}
\beta \equiv (\gamma_{00})^{1/2}
\ee
(in any coordinates $(y^\mu)$ adapted to the frame E, and such that $y^0 = cT$). Metric $\Mat{g}$ is related to the Euclidean metric $\Mat{g}^0$ by the assumed gravitational contraction of objects, including space standards---hence the dilation of measured distances. This effect was formerly assumed to occur only in the direction of $\mathrm{grad}\,\rho_e$ (see e.g. Ref. \cite{A34}), but it is now assumed isotropic. This means that the following relation is now assumed \cite{O3} between the spatial metrics $\Mat{g}^0$ and $\Mat{g}$: 
\begin{equation} \label {spacemetric}
		 			\Mat{g} =	\beta^{-2}\Mat{g}^0.\\
\end{equation}

\subsection{Dynamical equations} \label{Dynamics}
In addition to its metrical effects, the gravitational field $p_e$ has also direct dynamical effects, namely it produces a gravity acceleration
\be \label{vecteur_g}
\mathbf{g} \equiv -\frac{\mathrm{grad}_{\Mat{g}}\, p_e}{ \rho_e}.
\ee
[The index $_{\Mat{g}}$ means that the gradient operator refers to the physical, Riemannian metric $\Mat{g}$, {\it i.e.}, $(\mathrm{grad}_{\Mat{g}}\phi)({\bf x})$ is a spatial vector (an element of TM$_{\bf x}$) with components $(\mathrm{grad}_{\Mat{g}}\phi)^i=g^{ij}\phi_{,j}$.] We shall soon assume that the $p_e$-$\rho_e$ relationship is simply $p_e=c^2\rho _e$. In that case, we get from (\ref{beta}):
\be \label{vecteur_g_beta}
\mathbf{g} = -c^2\frac{\mathrm{grad}_{\Mat{g}}\,\beta }{\beta }.
\ee
Equation (\ref{vecteur_g}) is a fundamental one according to the concept of the theory. However, to use this ``gravity acceleration," we must define the dynamics by an extension of Newton's second law, as announced in the Introduction. This has been done in detail in Refs. \cite{A16,A15,A20} (see Ref. \cite{A34} for a summary). {\it The dynamical equations stay unchanged in the new version of the theory}. Therefore, only a synopsis of that alternative dynamics will be given here. Dynamics of a test particle is governed by a ``relativistic" extension of Newton's second law:
\begin{equation} \label{Newtonlawmasspoint} 
\mathbf{F}_0 + m(v)\mathbf{g} = \frac{D\mathbf{P}}{Dt_\mathbf{x}},		 		 
\end{equation}
where $\mathbf{F}_0$ is the non-gravitational ({\it e.g.} electromagnetic) force, $\mathbf{v}\equiv d\mathbf{x}/dt_{\mathbf{x}}$ the velocity and $v\equiv \Mat{g}(\mathbf{v},\mathbf{v})^{1/2}$ its modulus, both being measured by physical clocks and rods of observers bound to the preferred frame E; $m(v) \equiv m(0)\gamma_v$ is the relativistic mass  ($\gamma_v\equiv (1-v^2/c^2)^{-1/2}$ is the Lorentz factor); $\mathbf{P} \equiv m(v) \mathbf{v}$ is the momentum; $t_\mathbf{x}$ is the ``local time" in $\mathrm{E}$, measured by a clock at the fixed point $\mathbf{x}$ in the frame $\mathrm{E}$, that momentarily coincides with the position of the test particle: from (\ref{spacetimemetric}), we have
\begin{equation} \label {localtime}
		 			dt_\mathbf{x}/dT = \beta(T,\mathbf{x});
\end{equation}
and $D\mathbf{w}/D\xi$ is the appropriate ``time-"derivative of a vector $\mathbf{w}(\xi)$ in a manifold $\mathrm{M}$ endowed with a ``time-"dependent metric $\Mat{g}_\xi $ \cite{A16,A15}. In the static case ($\beta_{,0}=0$) with $\mathbf{F}_0 = {\bf 0}$, that dynamics implies Einstein's motion along geodesics of the curved space-time metric $\Mat{\gamma}$ (see Ref.  \cite{A16}). 
\footnote{\,
$\beta$ being defined by Eq. (\ref{beta_operatoire}), postulating Eq. (\ref{vecteur_g_beta}) for vector ${\bf g}$ is {\it equivalent}, under natural requirements, to asking geodesic motion in the static case \cite{A16}. On the other hand, if one adds to the r.h.s. of Eq. (\ref{vecteur_g_beta}) a complementary term involving the velocity of the test particle and the time-variation of the space metric, then Eq. (\ref{Newtonlawmasspoint}) implies geodesic motion in the general case, thus one deduces Einstein's geodesic motion from a relativistic extension of Newton's second law \cite{A16}.
} 
\\

Dynamics is also defined for continuous media. For a {\it dust}, we may apply (\ref{Newtonlawmasspoint}) pointwise and this implies \cite{A20} the following equation: 
\begin{equation} \label{continuum}
T_{\mu;\nu}^{\nu} = b_{\mu},				          
\end{equation}
where ${\bf T}$ is the energy-momentum tensor of matter and nongravitational fields, and where $b_\mu$ is defined by
\begin{equation} \label{definition_b}
b_0(\mathbf{T}) \equiv \frac{1}{2}\,g_{jk,0}\,T^{jk}, \quad b_i(\mathbf{T}) \equiv -\frac{1}{2}\,g_{ik,0}\,T^{0k}. 
\end{equation}
(Indices are raised and lowered with metric $\Mat{\gamma}$, unless mentioned otherwise. Semicolon means covariant differentiation using the Christoffel connection associated with metric $\Mat{\gamma}$. Note that in GR, in contrast, we have $b_\mu=0$ in (\ref{continuum}).) The universality of gravity means that Eq. (\ref{continuum}) with definition (\ref{definition_b}) must remain true for any material medium. {\it Thus, the dynamics of a test particle, as well as the dynamical field equation for a continuous medium, exactly obey the WEP, just as it is the case in GR.} Equations (\ref{continuum})-(\ref{definition_b}) are valid in any coordinates $(y^\mu)$ which are adapted to the frame E and such that $y^0=\phi(T)$ for some function $\phi$. 

\section{Scalar field equation and energy conservation} \label{field+energy} 
\subsection{Semi-heuristic constraints on the scalar field equation}
Equation (\ref{vecteur_g}) for the gravity acceleration expresses the idea according to which gravity would be Archimedes' thrust in Romani's ``constitutive ether" \cite{Romani}, {\it i.e.}  a space-filling perfect fluid, of which any matter particle should be a mere local organization. The equation for the field $p_e$ in (\ref{vecteur_g}) should fulfil the following conditions \cite{O3}:\\
 {\bf i}) Newton's gravity, because it propagates instantaneously, should correspond to the limiting case of an incompressible ether ($\rho _e = \mathrm{Constant}$). \\
 {\bf ii}) In the real case, the ether should have a compressibility $K = 1/c^2$, so that the velocity of the pressure waves, $c_e \equiv (\dd p_e/\dd\rho_e)^{1/2}$ (beyond which velocity the material particles, seen as flows in the universal fluid, should be destroyed by shock waves), coincide with the relativistic upper limit $c$. (This constraint leads to $p_e=c^2 \rho_e$, as assumed in advance for Eq. (\ref{vecteur_g_beta}).) Thus, in the real case, the equation for $p_e$ should be a kind of nonlinear wave equation, the nonlinearity arising from the fact that the physical space-time metric $\Mat{\gamma}$ is determined by the field $p_e$ itself. It is indeed the physical metric, not the background metric, which is directly relevant here, because the relativistic upper limit $c$ applies to velocities measured with the local, physical instruments. \\

We note that, from Eqs. (\ref{spacetimemetric}) and (\ref{spacemetric}), the physical metric $\Mat{\gamma }$ is nearly equal to the given Galilean metric $\Mat{\gamma }^0$ if and only if $\beta \approx 1$, which, by the definition (\ref{beta}), means a quasi-incompressible flow, $\rho_e \approx \mathrm{Constant}$ if moreover we consider a ``non-cosmological time scale" so that $\rho_e^\infty \approx \mathrm{Constant}$. (Due to Eq. (\ref{vecteur_g_beta}), it also implies that the gravitational field is weak, {\it i.e.} the gravity acceleration is ``small.") Thus, in the limit $\beta \rightarrow 1$ with $\rho_e^\infty \approx \mathrm{Constant}$, the nonlinearity of the wave equation must evanesce while simultaneously the ether compressibility becomes very small and the metric becomes close to Galilean. Now Newton's gravity does correspond to a Galilean metric and is characterized by Poisson's equation for the field ${\bf g}$:
\be \label{Poisson_g}
\mathrm{div}_{\Mat{g}^0} \ {\bf g}= - 4 \pi G \rho, 
\ee
with $G$ the gravitational constant and $\rho $ the Newtonian mass density. Hence, for an incompressible ether ($\rho_e = \mathrm{Constant}$), in which the gravity acceleration is defined by Eq. (\ref{vecteur_g}) with $\Mat{g}=\Mat{g}^0$, Newton's gravity is exactly equivalent to the following equation for the ``ether pressure" $p_e$:
\be \label{Poisson_p_e}
\Delta_{ \Mat{g}^0 }\, p_e =  4 \pi G \rho \rho_e.\\
\ee

Therefore, the conditions {\bf i}) and {\bf ii}) suggest to admit that the equation for the field $p_e$ has the form \cite{O3}
\be \label{possibleField}
\Delta_{\Mat{g}}\, p_e + (\mathrm{time\ derivatives}) =  4 \pi G \sigma \rho_e F(\beta), 
\ee
\be \label{limitF=1}
F(\beta) \rightarrow  1  \mathrm{\ as\ }  \beta \rightarrow 1,                   
\ee
the ``time derivatives" term being such that, in the appropriate (``post-Minkowskian") limit \cite{A34}, involving the condition $\beta \rightarrow 1$, the operator on the l.h.s. becomes equivalent to the usual (d'Alembert) wave operator. On the r.h.s. of (\ref{possibleField}), $\sigma$ denotes the relevant mass-energy density, which shall have to be defined in terms of the energy-momentum tensor $\mathbf{T}$ (that we take in mass units), and which will reduce to the Newtonian (rest-mass) density $\rho$ in the nonrelativistic limit. It is worth noting that the equation assumed \cite{A34}  for the previous (anisotropic-metric) version of the theory is indeed a special case of Eq. (\ref{possibleField}) \cite{O3}. Let us thus start from (\ref{possibleField}) and (\ref{limitF=1}), or equivalently from (\ref{limitF=1}) and
\be \label{possibleFieldrho}
\Delta_{\Mat{g}}\, \rho _e + (\mathrm{time\ derivatives}) =  \frac{4 \pi G}{c^2} \sigma \rho_e F(\beta).
\ee

\subsection{The equation for the scalar gravitational field and the energy equation}

As we mentioned, the physical metric is directly relevant to translate the heuristic considerations on the ``gravitational ether" into restrictions imposed on the field equation. However, the latter would obviously be more tractable if it could be rewritten in terms of the flat background metric. Let us try to impose this condition. To this aim, we evaluate the spatial term $\Delta_{\Mat{g}}\, \rho _e$ on the l.h.s. of (\ref{possibleFieldrho}). From the definition
\be \label{Delta_g}
\Delta_{\Mat{g}}\, \phi \equiv \mathrm{div}_{\Mat{g}} (\mathrm{grad}_{\Mat{g}} \phi ) = \frac{1}{\sqrt{g}} \left(\sqrt{g}\, g^{ij}\phi_{,j}\right)_{,i} \qquad \left(g \equiv \mathrm{det}(g_{ij}), \quad (g^{ij}) \equiv (g_{ij})^{-1}\right),
\ee
and since, by (\ref{spacemetric}), we have 
\be \label{det_g}
g = \beta^{-6} g^0 \qquad \left(g^0 \equiv \mathrm{det}(g^0_{ij})\right),
\ee
it follows that, in Cartesian coordinates for $\Mat{g}^0$, it holds
\be \label{Delta_rho_e}
\Delta_{\Mat{g}}\, \rho_e = \beta^3\left(\frac{\rho_{e,i}}{\beta }\right)_{,i}.
\ee
Accounting for the definition of $\beta$ from $\rho_e$ (Eq. (\ref{beta})), we get:
\be \label{DeltaLogbeta}
\Delta_{\Mat{g}}\, \rho_e = \rho_e^\infty\, \beta^3 \left(\frac{\beta_{,i}}{\beta }\right)_{,i}=\rho_e^\infty \, \beta^3 \, \Delta_{\Mat{g}^0} \left(\mathrm{Log}\beta\right).
\ee
Hence, if the sought-for finalization of Eq. (\ref{possibleFieldrho}) is to be reexpressed nicely in terms of the flat metric, the field variable should be 
\be \label{def_psi}
\psi \equiv - \mathrm{Log}\beta.
\ee
(The minus sign is chosen so that $\psi \geq 0$, see Eq. (\ref{beta}).)\\

On the other hand, we impose on the field equation the additional condition that an {\it exact local conservation law} must be found with some consistent definition of the energy. The latter should be the sum of a material energy and a gravitational energy. The energy conservation law should be obtained by rewriting the time component of the dynamical equation, (\ref{continuum}) with the definition (\ref{definition_b}), as a zero 4-divergence. For any metric having the form (\ref{spacetimemetric}), and independently of any restriction on the equation for the scalar field, the following identity:
\be 
T_{ \mu ;\nu}^{\nu } = \frac{1} {\sqrt{-\gamma}} \left(\sqrt{-\gamma}\, T_\mu^{\nu }\right)_{,\nu } - \frac{1}{2}\gamma _{\lambda \nu ,\mu } T^{\lambda \nu } \qquad \left(\gamma  \equiv \mathrm{det}(\gamma _{\mu \nu })\right) 
\ee
allows us to rewrite the time component of (\ref{continuum}) as
\be \label{alpha}
\left(\sqrt{-\gamma}\, T_0^j\right)_{,j } + \left(\sqrt{-\gamma}\, T_0^0\right)_{,0} = \sqrt{-\gamma}\,\beta \beta_{,0}T^{00} \equiv \alpha.
\ee
Thus, $\alpha $ must be a 4-divergence by virtue of the equation for the scalar field. Using now the specific form (\ref{spacemetric}) assumed for the space metric, we have Eq. (\ref{det_g}) and get from (\ref{spacetimemetric}):
\be \label{gamma}
-\gamma = \beta^2 g = \beta^{-4}\,g^0,
\ee
so that, in Cartesian coordinates for $\Mat{g}^0$ and with $x^0 =cT$,
\be \label{alpha_isotropic}
\alpha = \left(\mathrm{Log}\,\beta\right)_{,0}T^{00}\equiv -\psi _{,0}T^{00}.
\ee
Together with (\ref{DeltaLogbeta}), Eq. (\ref{alpha_isotropic}) makes it obvious that the relevant scalar field is indeed $\psi \equiv - \mathrm{Log}\beta$. We note the identities
\be \label{psi0psijj}
\psi _{,0} \,\psi _{,j,j} = \left(\psi _{,0}\,\psi _{,j}\right)_{,j}-{\textstyle\frac{1}{2}} \left(\psi _{,j}\psi _{,j}\right)_{,0}, \qquad \psi _{,0} \,\psi _{,0,0} = {\textstyle\frac{1}{2}} \left(\psi _{,0}^2\right)_{,0}. 
\ee
From these, and from (\ref{alpha_isotropic}), it follows that, if we take the gravitational source $\sigma$ to be the energy component $T^{00}$, and if we simply postulate for $\psi$ the flat wave equation:
\begin{equation}\label{field}
\square \psi  \equiv \psi _{,0,0}-\Delta_{\Mat{g}^0}\, \psi  = \frac{4 \pi G}{c^2} \sigma \qquad (\sigma \equiv T^{00}, \quad x^0 \equiv cT),			  
\end{equation}
then we indeed obtain $\alpha$ as a 4-divergence:
\be \label{alpha=4div}
\alpha = \frac{c^2}{4\pi G}\left\{- {\textstyle\frac{1}{2}} \partial_0\left[\psi_{,0}^2+\left(\mathrm{grad}\psi \right)^2\right]+\mathrm{div} \left(\psi_{,0}\mathrm{grad}\psi \right)\right\}.
\ee
(Henceforth, div, grad and also $\Delta$ shall be the standard operators defined with the Euclidean metric $\Mat{g}^0$.) More complicated time derivatives in the field equation could also provide a conservation equation, but the spatial term is more or less imposed to be $\Delta \psi$ by (\ref{DeltaLogbeta}), while the source $\sigma$ should be $T^{00}$ due to (\ref{alpha_isotropic}). We have currently few constraints on the time-derivative part of the equation for the scalar gravitational field. Hence, {\it Occam's razor leads us to state (\ref{field}) as the equation for the scalar field.} (This corresponds to $F(\beta)=\beta^2$ in Eq. (\ref{possibleFieldrho}).)\\

Thus, by assuming the validity of Eq. (\ref{field}), we rewrite $\alpha $ [Eq. (\ref{alpha})] as (\ref{alpha=4div}). Hence, (\ref{alpha}) becomes the following {\it local conservation equation for the energy:}
\footnote{\,
We use the fact that, from (\ref{spacetimemetric}) and (\ref{gamma}), $\sqrt{-\gamma}\, T_0^0 = \beta^{-2}T_0^0 = T^{00}$ and $\sqrt{-\gamma}\, T_0^j= \beta^{-2}T_0^j= T^{0j}$ in Cartesian coordinates, so that (\ref{energyconservation}) with (\ref{energydensity}) and (\ref{energyflux}) apply then---but these are space-covariant equations.
}
\begin{equation} \label{energyconservation}
\partial_0(\varepsilon_\mathrm{m} + \varepsilon_\mathrm{g})+\mathrm{div}({\bf \Phi}_\mathrm{m}+ {\bf \Phi} _\mathrm{g}) = 0, 
\end{equation}
where the material and gravitational energy densities are given (in mass units) by:
\be \label{energydensity}
\varepsilon_\mathrm{m} \equiv T^{00}, \qquad \varepsilon_\mathrm{g} \equiv \frac{c^2}{8\pi G} \left[\psi_{,0}^2+\left(\mathrm{grad}\psi \right)^2\right],
\ee
and the corresponding fluxes are:
\be \label{energyflux}
 {\bf \Phi}_\mathrm{m}\equiv  (T^{0j}), \qquad {\bf \Phi}_\mathrm{g} \equiv -\frac{c^2}{4\pi G}  \left(\psi_{,0}\mathrm{grad}\psi \right).
\ee
The scalar field equation (\ref{field}) and the conservation equation (\ref{energyconservation}) are valid in any coordinates $(y^\mu)$ adapted to the preferred frame and such that $y^0=x^0\equiv cT$. We note in particular that, although the d'Alembert operator $\square\,$ is generally-covariant, Eq. (\ref{field}) does not admit a change in the time coordinate $y^0$, because $\psi \equiv - \mathrm{Log}\sqrt{\gamma _{00}}$ and $\sigma \equiv T^{00}$ behave differently under a change $y'^0=\phi(y^0)$.
\footnote{\,
However, we note also that a mere change in the time unit, $T'=aT$, does not affect the time coordinate $x^0\equiv cT$ (since $c$ becomes $c'=c/a$), hence leaves Eqs. (\ref{field}) and (\ref{energyconservation}) invariant.
}
Thus, Eq. (\ref{field}) admits only purely spatial coordinate changes, consistently with the preferred-frame character of the theory. (It is recalled at $\S$ \ref{light-rays} how to cope with this character, on the example of the effects on light rays; see at the end of Ref. \cite{B21} for the case with celestial mechanics.) 

\subsection{Comments on the balance equation for the spatial momentum}

Similarly, let us rewrite the spatial component of the equation of motion of a continuum (\ref{continuum}) in terms of the scalar field (\ref{def_psi}), using the explicit form (\ref{spacetimemetric})-(\ref{spacemetric}) of the metric, and the identity
\be \label{T^munu_;nu}
T^{\mu \nu}_{;\nu } = \frac{1}{\sqrt{-\gamma }}\left(\sqrt{-\gamma }\,T^{\mu \nu }\right)_{,\nu }+ \Gamma '^\mu _{\nu \lambda }T^{\nu \lambda }
\ee
(where the $\Gamma '^\mu _{\nu \lambda }$'s are the Christoffel symbols of metric $\Mat{\gamma }$). {\it Adopting Galilean coordinates $(x^\mu)$ for the flat metric $\Mat{\gamma}^0$ henceforth}, we find after an easy algebra:
\be \label{space_continuum}
\left(e^{4\psi}\,T^{ij}\right)_{,j} + \left(e^{4\psi}\,T^{i0}\right)_{,0} - e^{4\psi} \left(\psi _{,i}T^{jj}+\psi _{,0}T^{i0}\right)= \psi _{,i} \sigma .
\ee
An identity similar to (\ref{psi0psijj}) (with $\psi _{,i}$ instead of $\psi _{,0}$) allows to get the r.h.s. as a 4-divergence, using the scalar field equation (\ref{field}). Due to the remaining source term on the l.h.s., it is in general not possible to rewrite (\ref{space_continuum}) as a zero flat 4-divergence. {\it I.e.}, there is no local conservation equation for the total (material plus gravitational) momentum in this theory. \\

In contrast, in Lagrangian-based relativistic theories of gravitation, e.g. in GR, there is a local conservation law (or something that looks like that) for the {\it total} momentum, which is the sum of the local momentum of {\it matter} and the local (pseudo-)momentum ${\bf \Theta }$ of the {\it gravitational field}. (The meaning of the latter decomposition and of its coordinate dependence is clearer in the teleparallel equivalent of GR \cite{Maluf-Faria}.) In some cases, characterized by a sufficient fall-off of ${\bf \Theta }$ at spatial infinity, the global value (i.e. the space integral) of the {\it total} momentum is then conserved. 
\footnote{\ 
In fact, it does not seem completely clear what should be the physically motivated conditions ensuring the sufficient decrease at infinity for ${\bf \Theta }$, due to the fact that one has to account for gravitational radiation: see e.g. Stephani \cite{Stephani}.
} 
However, this does not mean that the global value of the momentum of {\it matter} is then conserved: in fact, it {\it precludes} this, unless the global momentum of the gravitational field is separately constant---but this occurs only when the gravitational field is constant, thus when there is no motion of matter. Therefore, the situation is not so much different in the investigated theory and in Lagrangian relativistic theories: in both kinds of theories, the global momentum of {\it matter} is in general not conserved, unless there is just one body in equilibrium---the latter case is, of course, possible also in the investigated theory. [Assume time-independent fields in Eqs.~(\ref{field}), (\ref{energyconservation}) and (\ref{space_continuum}).] In Newton's theory, in which there is a local conservation for the total momentum, the global momentum of matter is conserved, however. This is because the global value of the gravitational momentum turns out to be {\it zero} in Newton's theory \cite{A15}. (The physical reason for this is that there is no gravitational radiation in Newton's theory.) The fact that, in contrast, the momentum of matter is in general not conserved in relativistic theories of gravitation, is related to the generic presence of self-accelerations (or self-forces) in these theories (including GR), already mentioned in the Introduction. 

\section{Equations of motion and matter production for a perfect fluid } \label{fluid} 

In most applications of a ``relativistic" theory of gravitation, it is enough (and it is indeed usual \cite{Fock59, Chandra65, Weinberg, MTW, Will93}) to consider a perfect fluid, because: {\bf i}) the stress tensor, {\it i.e.}, the spatial part of tensor ${\bf T}$, has normally a non-spherical part small enough that the latter does not bring significant post-Newtonian (PN) corrections; and {\bf ii}) the motion of astronomical bodies can be described as approximately rigid (here also, it is an even better approximation if one assumes this only at the stage of calculating the PN {\it corrections}), in which case a viscosity has no effect. For a perfect fluid, with its well-known expression for tensor ${\bf T}$ \cite{Fock59}, depending on the pressure $p$, the proper density of rest mass $\rho ^\ast$, the density of elastic energy per unit rest mass $\Pi$, and the velocity ${\bf u} \equiv  \dd {\bf x}/\dd T =\beta {\bf v}$, we introduce the field variable
\be \label{def_theta}
\theta \equiv e^{4\psi}\left(\sigma + {\textstyle\frac{p}{c^2}}\,e^{2\psi} \right), \qquad \sigma \equiv T^{00} = \left[\rho^\ast \left(1 + \frac{\Pi}{c^2}\right)+\frac{p}{c^2}\right] \frac{\gamma_v^2}{\beta^2} - \frac{p}{c^2\,\beta^2}
\ee
and rewrite the equation of motion (\ref{space_continuum}) and the energy conservation (\ref{energyconservation}) respectively as:
\be \label{spacefluid}
\left(\theta u^i \right)_{,T} + \left(\theta u^i u^j \right)_{,j} -\psi _{,T}\,\theta u^i - \psi _{,i}\, \theta u^j u^j = c^2 \psi _{,i}\, \sigma + e^{2\psi}\left(p\,\psi _{,i} -p_{,i} \right)
\ee
and
\be \label{timefluid}
\left(e^{-4\psi} \theta \right)_{,T} + \left(e^{-4\psi} \theta u^j \right)_{,j} = -\psi _{,T}\,\sigma +\frac{1}{c^2} \left(e^{2\psi} p \right)_{,T}.\\
\ee
\newline

As it has been discussed in detail in Ref. \cite{A20}, the exact energy conservation of the scalar theory precludes in general an exact conservation of (rest-){\it mass}. There, it has been shown that, already for a perfect and {\it isentropic} fluid, the general form (\ref{continuum}) of the equations of motion for a continuum implies a {\it reversible creation/destruction of matter in a variable gravitational field}. Let us note ${\sf U}$ the 4-velocity, with $U^\mu \equiv \dd y^\mu/\dd s$. One finds that the rate of creation/destruction is \cite{A20}: 
\be \label{creationrate}
\hat\rho \equiv \left(\rho ^\ast  U^\mu \right)_{;\mu} = \frac{pU^0}{2c^2}\Phi /\left(1+\frac{\Pi + p/\rho^\ast }{c^2} \right), \qquad \Phi \equiv \frac{g_{,0}}{g}.
\ee
This equation holds true independently of the scalar field equation and the specific form of the space metric \cite{A20}. With the new form (\ref{spacemetric}) assumed for the space metric, we get:
\be \label{Phi}
\Phi = -6 \frac{\beta_{,0}}{\beta } ,
\ee
which is three times the rate found with the formerly-assumed anisotropic space metric \cite{A20}. Of course, the actual values of $\beta$ and $\beta _{,0}$ in a given physical situation may depend on the theory. However, anticipating over the next Section, we can write down the weak-field approximations of $\beta$ and its relative rate as:
\be \label{beta_weakfield}
\beta \approx 1-\frac{U}{c^2}, \qquad \frac{\beta_{,0}}{\beta } \approx \frac{1}{c^3} \frac{\partial U}{\partial T},
\ee
with $U$ the Newtonian potential, whose time-derivative has to be taken {\it in the preferred frame}. This, indeed, gives three times the former weak-field prediction for $\hat\rho$, namely it gives
\be \label{creation_weakfield}
\hat\rho \approx \frac{3p}{c^4} \frac{\partial U}{\partial T},
\ee
which remains {\it extremely} small in usual conditions (but would be significant inside stars): the relative creation rate $\hat\rho/\rho$ would be ca. $10^{-22}\, \mathrm{s}^{-1}$ at or near the surface of the Earth, if the ``absolute" velocity of the Earth is taken to be $300\, \mathrm{km.s}^{-1}$. \{See Ref. \cite{A20}, Sect. 4.3. Note that equal amounts of mass would be destroyed and created at opposite positions on the Earth, Eq.~(4.22) there.\} Mass conservation is far to have been checked to this accuracy. Note that matter creation is being actively investigated in cosmological literature, see e.g. \cite{Prigogine89,Calvao92,Sussman94,Al-Rawaf96,Singh2000,BorgesCarneiro2005}. If mass non-conservation is to occur in ``cosmological" conditions, it must exist in nature, and so possibly (in minute, not yet observable quantities) in today's world.
\section{Post-Newtonian approximation (PNA)} \label{PNA} 
\subsection{Definition of the asymptotic scheme of PNA} \label{Principle_PNA}

The purpose of the post-Newtonian approximation is to obtain {\it asymptotic expansions} of the fields as functions of a relevant field-strength parameter $\lambda$, and to deduce expanded equations (which are much more tractable than the original equations) by inserting the expansions into the field equations. To do this in a mathematically meaningful way, it is necessary that one can make $\lambda $ tend towards zero, hence one must (conceptually) associate to the given gravitating system S a {\it family} $(\mathrm{S}_\lambda)$ of systems, {\it i.e.}, a family of solution fields of the system of equations. This family has to be defined by a family of boundary conditions---initial conditions for that matter, because here gravitation propagates with a finite velocity. The system of interest, S, must itself correspond to a small value $\lambda _0$ of the parameter, thus ``justifying" to use the asymptotic expansions for that value. The definition of the family involves two conditions which make this family represent the Newtonian limit: as $\lambda \rightarrow  0$, {\bf i}) the physical metric $\Mat{\gamma }^{(\lambda)}$ must tend towards the flat metric $\Mat{\gamma }^0$, and {\bf ii}) the fields must become equivalent to ``corresponding" Newtonian fields. The first condition is easy to be explicited in a scalar theory, in which the relation between $\Mat{\gamma }$ and $\Mat{\gamma }^0$ depends only on the scalar gravitational field. Condition {\bf ii}) asks for two preliminaries: {\it a}) that one disposes of a relevant family of {\it Newtonian} systems, for comparison, and {\it b}) that one is able to define a natural equivalent of the Newtonian gravitational field, {\it i.e.}, the Newtonian potential $U_{\mathbf{N}}$. (The matter fields for a perfect fluid are the same in a ``relativistic" theory as in Newtonian gravity (NG), up to slight modifications.) Point {\it a}) is easily fulfilled, once it is recognized \cite{FutaSchutz,A23} that there is an exact similarity transformation in NG, which is appropriate to describe the weak-field limit in NG itself \cite{A23}. This immediately suggests defining the family  $(\mathrm{S}_\lambda)$ {\it by applying the similarity transformation of NG to the initial data} defining a gravitating system in the investigated theory \cite{FutaSchutz,A23}---preferably to that initial data, of a general-enough nature, which precisely defines the system of interest, S \cite{A23}. Of course, to use the Newtonian transformation demands that point {\it b}) has been solved, which depends on the precise equations of the theory.\\

The application of these principles to the scalar ether-theory has been done in detail for its first version \cite{A23}. (See Ref. \cite{A25-26}, Sect. 2, for a synopsis and a few complementary points.) The modifications to be done for the present version are straightforward. The definition of the metric (\ref{spacetimemetric})-(\ref{spacemetric}) and that of the scalar field $\psi$ (\ref{def_psi}) imply that condition {\bf i}) is equivalent to asking that $\psi^{(\lambda)} \rightarrow 0$ as $\lambda \rightarrow 0$. Therefore, we may define a dimensionless weak-field parameter simply as
\be \label{def_lambda}
\lambda \equiv \mathrm{Sup}_{{\bf x} \in \mathrm{M}}\, \psi({\bf x})
\ee
(at the initial time, say). Moreover, from the scalar field equation (\ref{field}), we see that 
\be \label{def_V}
V \equiv c^2 \psi
\ee 
satisfies the wave equation with the same r.h.s. (in the Newtonian limit where $\sigma \sim \rho$) as Poisson's equation of NG, and the retardation effects should become negligible in the Newtonian limit. Hence, $V$ is a natural equivalent of $U_\mathrm{N}$. Thus, the Newtonian limit is defined by the same family of initial conditions as in the first version \cite{A23}, though with the new definition (\ref{def_V}) of $V$: at the initial time,
\footnote{\,
Since the given system is assumed to correspond to a small value $\lambda _0\ll 1$, the transformation goes first from $\lambda_0$ to $\lambda =1$, and then from $\lambda =1$ to the arbitrary value $\lambda$. This amounts to substituting $\xi \equiv \lambda /\lambda _0$ for $\lambda $, and $p^{(\lambda _0)}$, etc., for $p^{(1)}$, etc. \cite{A23}. Moreover, we consider a barotropic fluid: $\rho^\ast =F(p)$. Thus, the initial conditions for $p$ and for $\rho ^\ast$ are actually not independent, and one must assume that $F^{(\lambda)} (p)=\lambda F^{(1)}(\lambda ^{-2} p)$ \cite{A23}. Note that the small parameter $\lambda $ considered in the present paper corresponds to $\varepsilon^2$, where $\varepsilon$ is that used in Ref. \cite{A23}.
}
\begin{equation}\label{PN_IC1}
p^{(\lambda)}(\mathbf{x})=\lambda^2p^{(1)}(\mathbf{x}), \quad
\rho^{\ast (\lambda)}(\mathbf{x})=\lambda\rho^{\ast (1)}(\mathbf{x}),
\end{equation}
\begin{equation}\label{PN_IC2}
V^{(\lambda)}(\mathbf{x})=\lambda V^{(1)}(\mathbf{x}), \quad\partial_TV^{(\lambda)}(\mathbf{x})=\lambda^{3/2} \partial _TV^{(1)}(\mathbf{x}),
\end {equation}
\begin{equation}\label{PN_IC3}
\mathbf{u}^{(\lambda)}(\mathbf{x})= \sqrt {\lambda }\, \mathbf{u}^{(1)}(\mathbf{x}).
\end {equation}
The system $\mathrm{S}_\lambda$ is hence defined as the solution of the above initial-value problem $\mathrm{P}_\lambda$. One expects that the solution fields admit expansions in powers of $\lambda$, whose dominant terms have the same orders in $\lambda$ as the initial conditions. It is then easy to check that, by adopting $[\mathrm{M}]_\lambda = \lambda[\mathrm{M}]$ and $[\mathrm{T}]_\lambda = [\mathrm{T}]/\sqrt{\lambda}$  as the new units for the system $\mathrm{S}_\lambda$ (where $[\mathrm{M}]$ and $[\mathrm{T}]$ are the starting units of mass and time), {\it all fields become order $\lambda^0$, and the small parameter $\lambda$ is proportional to $1/c^2$} (indeed $\lambda=(c_0/c)^2$, where $c_0$ is the velocity of light in the starting units). Thus, the derivation of the 1PN expansions and expanded equations is very easy. At the first PNA, one writes first-order expansions in this parameter for the independent fields $V, p, {\bf u}$:
\be \label{expans_base}
V= V_0 + V_1/c^2 + O(c^{-4}),\qquad p= p_0 + p_1/c^2 + O(c^{-4}), \qquad {\bf u}= {\bf u}_0 + {\bf u}_1/c^2 + O(c^{-4}),
\ee
and one deduces expansions for the other fields. (Of course, all fields depend on the small parameter $\lambda \propto 1/c^2$.) In these varying units, we have $T = \lambda^{1/2}T_0$ where $T_0$ is the ``true" time, {\it i.e.}, that measured in fixed units. Hence $cT = c_0T_0$ is proportional to the true time. But since the true velocities in system $\mathrm{S}_\lambda$ vary like $\lambda^{1/2}$ [Eq. (\ref{PN_IC3})], it is $T$, not $cT \propto T_0$, which remains nearly the same, as $\lambda$ is varied, for one orbital period of a given body in the Newtonian limit. Therefore, in this limit, thus for PN expansions, one must take $x'^0 \equiv  T$ as the time variable, in the varying units utilized for the expansions. This means that, in these units, the expansions are first of all valid at fixed values of ${\bf x}$ and $T$; and one can differentiate them with respect to these variables, because it is reasonable to expect that the expansions are uniform in ${\bf x}$ taken in the ``near zone" occupied by the gravitating system, and in $T$ taken in an interval where the system remains quasi-periodic \cite{A23}. 

\subsection{Main expansions and expanded equations}

Inserting (\ref{expans_base})$_1$ into the scalar field equation (\ref{field}) and accounting for the fact that the time variable is $x'^0 \equiv  T$ (in the varying units utilized), yields after powers identification:
\be \label{expans_fieldeq}
\Delta V_0 = -4\pi G \sigma _0, \qquad \Delta V_1 = -4\pi G \sigma _1 + \partial_T^2 V_0,
\ee
where $\sigma = \sigma _0 + \sigma _1/c^2 +O(c^{-4})$ is the 1PN expansion of the active mass density. Thus, the retardation effect disappears in the PN expansions. (However, the ``propagating" (hyperbolic) character of the gravitational equations is maintained through the fact that an initial-value problem is considered.) From (\ref{expans_fieldeq}) with appropriate boundary conditions ($U=O(1/r)$ and $\mathrm{grad}\,U=O(1/r^2)$ as $r\rightarrow \infty$) \cite{A23}, it follows that $U \equiv V_0$ is the Newtonian potential associated with $\sigma _0$:
\be \label{U}
U \equiv V_0 = \mathrm{N.P.}[\sigma _0] \qquad \left(\mathrm{N.P.}[\tau](\mathbf{X},T) \equiv G \int  \tau (\mathbf{x},T)\dd {\mathsf V}(\mathbf{x})/\abs{\mathbf{X - x}}\right),
\ee
(${\mathsf V}$ will denote the Euclidean volume measure on the space M), and that (imposing the same boundary conditions to $B$ as those for $U$)
\be \label{V_1}
V_1 = B + \frac{\partial ^2 W}{\partial T^2}, \qquad B \equiv \mathrm{N.P.}[\sigma_1],
\ee
with
\begin{equation}\label{W}
  W(\mathbf{X},T) \equiv G \int \abs{\mathbf{X - x}}
 \sigma _0(\mathbf{x},T)\dd {\mathsf V}(\mathbf{x})/2.
\end{equation}
The mass centers will be defined as barycenters of $\rho $, the density of rest-mass in the preferred frame and with respect to the Euclidean volume measure ${\mathsf V}$ \cite{A25-26}. It is related to the proper rest-mass density $\rho ^\ast$ by Lorentz and gravitational contraction \cite{A15}, so that $\rho = \rho ^\ast \gamma_v \sqrt{g}/\sqrt{g^0}$, hence from (\ref{det_g}):
\be \label{rho_rho*}
\rho =  \rho ^\ast \gamma_v /\beta ^3.
\ee
Using this and the definitions (\ref{field})$_2$ and (\ref{def_theta}), and since we have from (\ref{U}): 
\be \label{expans_beta}
\beta =e^{-V/c^2} = 1-U/c^2 + O(c^{-4}),
\ee
we get:
\be \label{expans_matter_0}
\rho ^\ast _0=\rho _0 = \sigma _0 = \theta _0
\ee
and
\be \label{rho_1}
\rho_1 = \rho ^\ast_1 + \rho_0 \left(\frac{{\bf u}_0^2}{2} + 3U \right),
\ee
\be \label{sigma_1}
\sigma_1 = \rho_1 + \rho_0 \left(\frac{{\bf u}_0^2}{2} -U + \Pi_0 \right),
\ee
\be \label{theta_1}
\theta_1 = \sigma_1 + p_0 + 4 \rho_0 U =\rho_1 + \rho_0 \left(\frac{{\bf u}_0^2}{2} +3U + \Pi_0 \right) +p_0.\\
\ee
\newline

The expansion of the equation of motion (\ref{spacefluid}) and the energy equation (\ref{timefluid}) gives, at the order zero:
\begin{equation}\label{i-ord0}
    \partial _T (\rho_0 u_0^i) + \partial _j (\rho_0 u_0^i u_0^j)=\rho_0
    U_{,i}-p_{0,i},
\end{equation}
\begin{equation}\label{T-ord0}
  \partial _T \rho_0 + \partial _j (\rho_0 u_0^j) = 0,
\end{equation}
which are just the Newtonian equations. The expanded equations of the order one in $1/c^2$ are:
\begin{eqnarray}\label{i-ord1}
    \partial _T (\rho_0 u_1^i + \theta_1 u_0^i) + \partial _j ( \rho_0 u_0^i u_1^j + \rho_0 u_1^i u_0^j +\theta_1  u_0^i u_0^j ) - \rho_0 (u_0^i \partial_T U + {\bf u}_0^2 \partial_i U) = & & \nonumber \\ = \sigma_1 U_{,i} + \rho_0 V_{1,i}+ p_0 U_{,i} -2U p_{0,i}-p_{1,i} & , &
\end{eqnarray}
\begin{equation}\label{T-ord1}
  \partial _T (w_0+\rho_1) + \partial _j [(w_0+p_0+\rho_1)u_0^j +\rho_0
  u_1^{j}]=-\rho_0\, \partial_T U,\qquad w_0 \equiv\rho_0(\frac{\mathbf{u}_0^2}{2}+\Pi_0 -U).
\end{equation}
Combining (\ref{i-ord0}), the continuity equation (\ref{T-ord0}),
and the 0-order expansion of the isentropy equation:
\begin{equation}\label{isentropy}
  \dd\Pi_0 = -p_0 \, \dd(1/\rho_0),
\end{equation}
one gets in a standard way the Newtonian energy equation:
\begin{equation}\label{energyN}
  \partial _T w_0 + \partial _j [(w_0+p_0)u_0^j]=-\rho_0 \, \partial_T U.
\end{equation}
Subtracting (\ref{energyN}) from (\ref{T-ord1}) gives us
\begin{equation}\label{mass-ord1}
  \partial _T \rho_1 + \partial _j(\rho_1 u_0^j + \rho_0 u_1^j)=0,
\end{equation}
which (together with (\ref{T-ord0})) means that mass is conserved at the first PNA of the scalar theory, also in this second version.

\subsection{Application: gravitational effects on light rays} \label{light-rays}

The effects of a gravitational field on an electromagnetic ray, seen as a ``photon" (a test particle with zero rest-mass), represent the most practically-important modification to NG. In this theory, they can be obtained by applying the extension (\ref{Newtonlawmasspoint}) of Newton's second law, in which ${\bf F}_0= {\bf 0}$ and the mass content of the energy $E = h\nu$ has to be substituted for the inertial mass $m(v)$. In the new version of the scalar theory, things go in close parallel with the former version, based on an anisotropic space contraction \cite{A19,A34}:
\begin {itemize}

\item {\bf i}) The main step is the recognition \cite{A19} that the PN equation of motion for a photon, obtained thus, coincides with the PN expansion of the geodesic equation for a light-like particle in the space-time metric $\Mat{\gamma }$, because the $\Gamma '^i_{0j}={\textstyle\frac{1}{2}}g^{ik}\partial_0 g_{kj}$ Christoffel symbols of $\Mat{\gamma }$ are $O(c^{-2})$ and the $\Gamma '^{\,0}_{jk}={\textstyle\frac{1}{2}}\gamma^{00} \partial_0 g_{jk}$ are $O(c^{-4})$ (with $x'^0 \equiv T$ as the time coordinate). This holds true in the present version based on Eq. (\ref{spacemetric}) for the space metric, because the same expansion [Eq. (\ref{expans_beta}) above] applies as in the former version. Therefore, to compute the effects of a weak gravitational field on light rays, one has to study the PN expansion of $\Mat{\gamma }$ in the relevant reference frame: that frame $\mathrm{E}_{\bf V}$ which moves with the velocity ${\bf V}$ in the preferred frame, assumed constant and small as compared with $c$, of the mass-center of the gravitating system. In coordinates $(x'^\mu )$ that are Galilean for the flat metric $\Mat{\gamma }^0$ and adapted to the frame $\mathrm{E}_{\bf V}$, the components $\gamma '_{\mu \nu }$ of $\Mat{\gamma }$ are deduced from its components in the preferred frame (Eqs. (\ref{spacetimemetric}) and (\ref{spacemetric})) by a special Lorentz transform, relative to $\Mat{\gamma }^0$ \cite{A19}; hence the $\gamma '_{\mu \nu }$ 's depend only on the field $\beta$ (not any more on its derivatives, as was the case with the former version), and on the velocity $V$.  

\item {\bf ii}) Inserting the expansion (\ref{expans_beta}) of $\beta$, one gets the PN expansion of the $\gamma '_{\mu \nu }$ 's. The PN expansion of $\gamma'_{00}$ is enough to compute the gravitational redshift. It is still $\gamma'_{00}=1-2Uc^{-2}+O(c^{-4})$ with $U$ the Newtonian potential: this holds true in the present version (in particular, Eq. (52) of Ref. \cite{A19} holds true). To get the other two effects of gravitation on light rays, namely the deflection and the time delay, one needs to compute the PN expansion of all components $\gamma '_{\mu \nu }$. One finds easily that, as before \cite{A19}, $\gamma '_{0i}=O(c^{-3})$ (in fact $\gamma '_{0i}=0$ for $i=2$ and $3$, now); such $\gamma '_{0i}=O(c^{-3})$ component(s) have (has) no influence on the PN equation of motion of a photon (see the equation after Eq. (9.2.4) in Weinberg \cite{Weinberg}, and see Eqs. (9.1.16), (9.1.19) and (9.1.21) there). And one finds that $\gamma '_{ij}=-(1+2Uc^{-2})\delta _{ij}+O(c^{-4})$. Thus, in the new version of the scalar ether-theory, the PN equation of motion of a photon coincides with the PN geodesic equation of motion of a photon in the so-called \cite{Will93} ``standard PN metric" of GR, and this is true {\it also in the relevant frame} $\mathrm{E}_{\bf V}$. In particular, in the SSS case, the formulas for the PN effects on photons are the same as those derived from the (space-)isotropic form of Schwarzschild's metric---or also from the harmonic form of the Schwarzschild metric (which is the SSS solution of the RTG \cite{Logunov89}), since its PN approximation is space-isotropic \cite{Fock59} and coincides with the PN approximation of the isotropic form. These predictions are accurately confirmed by observations \cite{Will93}.

\end{itemize}
\section{PN equations of motion of the mass centers} \label{EMMC} 

As already mentioned, the mass centers are defined \cite{A25-26} as local barycenters of the density of rest-mass in the preferred frame, $\rho $ or rather $\rho _{\mathrm{exact}}$, Eq. (\ref{rho_rho*}). (Henceforth, the index 0 will be omitted for the zero-order (Newtonian) quantities, for conciseness; therefore, the exact quantities, when needed, are denoted by the index ``exact.") Integrating Eq. (\ref{i-ord1}) in the (time-dependent) domain $\mathrm{D}_a$ occupied by body $(a)$ ($a=1,..., N$) in the preferred frame E gives
\be \label{integ-i-ord1}
\frac{\dd}{\dd T} \left( \int_{\mathrm{D}_a} (\rho u_1^i + \theta _1 u^i) \dd {\mathsf V} \right) = \int_{\mathrm{D}_a} f_1^i \dd {\mathsf V},
\ee
with
\be \label{f1i}
f_1^i = (\sigma _1 + p)U_{,i} + \rho V_{1,i} - 2Up_{,i} + \rho (u^i \partial_T U + {\bf u}^2 U_{,i} ).
\ee
Accounting for Eq. (\ref{theta_1}) and for Eq. (3.21) of Ref. \cite{A25-26}, we get:  \begin{equation}\label{masscent-ord1}
  M_a^1\,\ddot{a}_1^i+\dot{I}^{ai}=J^{ai}+K^{ai},
\end{equation}
which is Eq. (4.9) of Ref. \cite{A25-26}, and with, as there,
\be \label{M_a^1}
M_a^1 \equiv \int_{\mathrm{D}_a} \rho _1  \dd {\mathsf V}, \qquad M_a^{1}\mathbf{a}_{1} \equiv \int_{\mathrm{D}_a}\rho_{1}\mathbf{x} \,\dd {\mathsf V}(\mathbf{x}),
\ee
but with modified definitions of $I^{ai}$, $J^{ai}$ and $K^{ai}$:
\begin{equation}\label{Iai}
  I^{ai}\equiv \int_{\mathrm{D}_a} [p+\rho(\mathbf{u}^2/2+\Pi+3U)]u^i \dd {\mathsf V},
\end{equation}
\begin{equation}\label{Jai}
  J^{ai}\equiv \int_{\mathrm{D}_a} (\sigma_1 U_{,i}+\rho V_{1,i}) \dd {\mathsf V},
\end{equation}
and
\begin{equation}\label{Kai}
  K^{ai}\equiv \int_{\mathrm{D}_a}[-2Up_{,i}+pU_{,i}+ \rho u^i\partial _TU + \rho {\bf u}^2 U_{,i}]\dd {\mathsf V} = \int_{\mathrm{D}_a}[3pU_{,i}+ \rho u^i\partial _TU + \rho {\bf u}^2 U_{,i}]\dd {\mathsf V} \equiv K_1^{ai} + K_2^{ai} + K_3^{ai}.
\end{equation}
\newline
Together with the Newtonian equation, Eq. (\ref{masscent-ord1}) allows in principle to calculate the 1PN motion of the mass centers: due to Eq. (3.15) of Ref. \cite{A25-26}, the 1PN acceleration of the mass center of body $(a)$ is given by
\be \label{1PN_accel}
{\bf A}^{a} \equiv {\bf \ddot{a}}_{(1)} =  {\bf \ddot{a}} + \frac{M_a^1({\bf \ddot{a}}_{1} -  {\bf \ddot{a}})}{c^2M_a},
\ee
in which $M_a$ and ${\bf \ddot{a}}$ are the Newtonian mass and acceleration. Equation (\ref{masscent-ord1}) may be made tractable for celestial mechanics, as was done in Ref. \cite{A32} for the former version of the theory, by taking benefit of: {\it a}) the good separation between bodies, and {\it b}) the fact that the main celestial bodies are nearly spherical. This is left to a future work. Here, we will study the point-particle limit of this equation and will show that the deadly violation of the WEP for a small body, which was found with the former version of the theory \cite{A33}, does not exist any more with the new version.

\section{Point-particle limit and the WEP} \label{pointlimit} 

In order to define that limit rigorously and generally, we consider (as in Ref. \cite{A33}) a family of 1PN systems, that are identical up to the size of the body numbered (1): this size is a small parameter $\xi$. We have to expand as $\xi \rightarrow 0$ the integrals (\ref{Iai}), (\ref{Jai}), and (\ref{Kai}), for the small body, {\it i.e.} $a=1$. (As to the zero-order (Newtonian) acceleration of the small body, it tends towards the acceleration of a test particle in the Newtonian field $U^{(a)}$ of the other bodies \cite{A33}, as expected.) To do that, we use the simplifying assumption according to which the Newtonian motion of the small body is a {\it rigid motion}. The calculations are very similar to those \cite{A33} with the previous version, though simpler for the $K^{ai}$ integral; hence, we shall be concise.

\subsection{The general case}

The modification of the calculations in Ref. \cite{A33}, Sect. 3, is immediate for $I^{ai}$ and $K^{ai}$. We get [reserving henceforth the letter $a$ for the first, small body (for which $a=1$ in fact) and using the letter $b$ for the other, massive bodies]:
\be\label{I-PP}
{\bf I}^a \equiv (I^{ai}) = M_a[{\textstyle\frac{1}{2}} {\bf \dot{a}}^2 + 3U^{(a)}({\bf a})]{\bf \dot{a}} + O(\xi^4),
\ee
\be\label{K1-PP}
{\bf K}_1^{a} = O(\xi^5), 
\ee
\be\label{K2-PP}
{\bf K}_2^{a} = GM_a {\bf \dot{a}}\sum_{b \ne a}  \frac{M_b{\bf (a-b).{\dot b}}}{\abs{{\bf a-b}}^3} +O(\xi^4), 
\ee
\be\label{K3-PP}
{\bf K}_3^{a} = M_a \, {\bf \dot{a}}^2 \nabla U^{(a)}({\bf a})  +O(\xi^4). 
\ee
\newline
As to the integral $J^{ai}$, it has the same expression as before \cite{A25-26}, but the PN correction $\sigma_1$ to the active mass density is now given by Eq. (\ref{sigma_1}). Therefore, the expansion (3.27) of Ref. \cite{A33} remains valid for $J^{ai}$, but we have now:
\be \label{alpha_a}
\alpha_a \equiv \int_{\mathrm{D}_a} \sigma _1 \dd {\mathsf V} = M_a^1 +M_a\,[{\textstyle\frac{1}{2}} {\bf \dot{a}}^2 -U^{(a)}({\bf a})]_{\mid T=0} +O(\xi^5), 
\ee
\be \label{M_a^1-PP}
M_a^1 = M_a\,[ {\textstyle\frac{1}{2}} {\bf \dot{a}}^2 +3U^{(a)}({\bf a})]_{\mid T=0} +O(\xi^5),
\ee
\be \label{beta_a}
\beta _{aj} \equiv \int_{\mathrm{D}_a} \sigma _1({\bf x})(x^j-a^j) \dd {\mathsf V}({\bf x}) = M_a^1(a_1^j-a^j) +O(\xi^4). 
\ee
Beside $\alpha_a$ and $\beta_{aj}$, Eq. (3.27) of Ref. \cite{A33} involves also all those multipoles of the densities $\rho $ and $\sigma _1$ that correspond to the {\it other} bodies. \\
Since all of these equations contain the (Newtonian) mass $M_a$ as a common factor, it follows that the 1PN acceleration ${\bf A}^{a} $ of the (mass center of the) small body $(a)$, Eqs.~(\ref{masscent-ord1}) and (\ref{1PN_accel}), does not depend on its mass $M_a$. It then also follows from these equations (including Eq. (3.27) of Ref. \cite{A33}) that, neglecting $O(\xi )$ terms in ${\bf A}^{a} $, it depends only: 
\begin{itemize}
\item on the current Newtonian positions and velocities of all bodies: ${\bf a}, {\bf b}, {\bf \dot{a}}, {\bf \dot{b}}$; 
\item on all current 1PN positions: ${\bf a}_{(1)}, {\bf b}_{(1)}$; 
\item on the Newtonian masses $M_b$ of the {\it other} bodies and on {\it their} Newtonian potential $U^{(a)}$ (the ``external" potential for $(a)$); 
\item and still, through the external multipoles of $\rho $ and $\sigma _1$, on the structure of the {\it other} bodies.
\end{itemize}
Thus, the 1PN acceleration of a freely-falling small body is independent of its mass, structure, and composition, in other words {\it the WEP is satisfied at the 1PN approximation with the new version of the scalar theory}, in contrast with what happened with the former version \cite{A33}. 

\subsection{Comparison with a test particle in the case with one SSS massive body}

Since the WEP is satisfied, it seems obvious that the 1PN acceleration of a small body should be equal, at the point-particle limit ($\xi \rightarrow 0$), to that of a test particle---and this in the general case. We check this in the particular case where, beside the small body $(1)$, there is just one massive body $(2)$, whose mass center stays fixed at the origin in the preferred frame, and whose Newtonian density $\rho$ is spherically symmetric.
\footnote{
Strictly speaking, body (2) is gravitationally influenced by the small body, hence it cannot stay exactly at rest in the preferred frame. However, the PN acceleration of the massive body (2), due to the small body (1), is $O(\xi^3)$ (Ref. \cite {A33}, Sect. 2), so that we may forget this influence for the present purpose.
}
We note
\be \label{notationSSS}
m \equiv M_1, \quad M \equiv M_2, \quad {\bf x} \equiv {\bf a}, \quad {\bf v} \equiv {\bf \dot{x}}, \quad r \equiv \abs{{\bf x}}, \quad {\bf n} \equiv {\bf x}/r, \quad {\bf x}_1 \equiv c^2({\bf a}_{(1)} - {\bf a}).
\ee
Adapting Sect. 4 of Ref. \cite{A33}, we find without difficulty:
\be \label{I1point-PP-SSS}
{\bf \dot{I}}^1 = -m\frac{GM}{r^2} \left[ \left(\frac{{\bf v}^2}{2} +3 \frac{GM}{r}\right) {\bf n} + 4 ({\bf v.n}){\bf v} \right],
\ee
\be \label{alpha_2-SSS}
\delta M \equiv \int_{\mathrm{D}_2} \sigma _1 \dd {\mathsf V}= \frac{17}{3} \varepsilon, \qquad \varepsilon \equiv \frac{1}{2}\int_{\mathrm{D}_2} \rho U \dd {\mathsf V},
\ee
\be \label{J1-PP-SSS}
{\bf J}^1 = m\frac{GM}{r^2} \left[ \left(-{\bf v}^2 +2 \frac{GM}{r} -4 \frac{GM}{r_0} -\frac{17}{3} \frac{\varepsilon }{M} \right) {\bf n} + \frac{1}{r} \left(3({\bf x}_1{\bf .n}){\bf n}-{\bf x}_1 \right) \right], \quad (r_0 \equiv r_{\mid T=0}),
\ee
\be \label{K1-PP-SSS}
{\bf K}^1 = -m\frac{GM}{r^2} {\bf v}^2 {\bf n}. 
\ee
Inserting these values into Eq.~(\ref{masscent-ord1}) and putting the result into (\ref{1PN_accel}), we get the equation for the 1PN correction ${\bf x}_1$ to the position of the mass center of the small body:
\be \label{PNcorr-PP-SSS}
{\bf \ddot{x}}_1 = \frac{GM}{r^2} \left[ \left(-{\bf v}^2 +4 \frac{GM}{r}  -\frac{17}{3} \frac{\varepsilon }{M} + 3\frac{{\bf x}_1{\bf .n}}{r} \right) {\bf n} + 4 ({\bf v.n}){\bf v} - \frac{{\bf x}_1 }{r} \right]. 
\ee
\newline
On the other hand, as recalled in Subsect.~\ref{Dynamics}, the equation of motion of a test particle in the scalar theory coincides, in the present static case, with the geodesic equation in the relevant metric---thus, here, the metric (\ref{spacetimemetric})-(\ref{spacemetric}), specialized to the SSS case, for which we get from (\ref{field}) and (\ref{def_psi}):
\be \label{beta-SSS}
\beta (T,{\bf X}) = e^{-\frac{GM'}{c^2 R}}, \quad (R \equiv \abs{{\bf X}} \geq r_2), 
\ee
with $r_2$ the radius of the spherical body (2), and with
\be \label{activemass-SSS}
M' \equiv  \int_{\mathrm{D}_2} \sigma _\mathrm{exact} \dd {\mathsf V} = M + \frac{\delta M}{c^2} +O(c^{-4}),
\ee
thus in Cartesian coordinates $(X^i)$ for the Euclidean metric $\Mat{g}^0$:
\be \label {SSSmetric}
\dd s^2 = \left[1-2\frac{GM'}{c^2 R}+2\left(\frac{GM'}{c^2 R}\right)^2 +O(c^{-6})\right] (\dd x^0)^2 - \left[1+2\frac{GM'}{c^2 R}+O(c^{-4})\right] \delta _{ij} \dd X^i \dd X^j,
\ee
which is the SSS form of the standard PN metric of GR. The corresponding (complete) equation of motion is given by Weinberg \cite{Weinberg} (Eq. (9.5.3) with here $\varepsilon =0$, $\Mat{\zeta } = {\bf 0}$, and $\phi =-GM'/r_{(1)}$). In our notation, this is
\be \label{PNmotion-testparticle-isotropic-SSS}
{\bf \ddot{x}}_{(1)} =  \frac{GM'}{r_{(1)}^2}\left\{ -{\bf n}_{(1)} +\frac{1}{c^2} \left[ \left(-{\bf v}_{(1)}^2 +4 \frac{GM'}{r_{(1)} }   \right) {\bf n}_{(1)} + 4 ({\bf v}_{(1)} .{\bf n}_{(1)}){\bf v}_{(1)}  \right] \right\},
\ee
with 
\be \label{PNnotationSSS}
{\bf x}_{(1)} \equiv {\bf x} + {\bf x}_1/c^2, \quad r_{(1)} \equiv \abs{{\bf x}_{(1)} }, \quad {\bf n}_{(1)}  \equiv {\bf x}_{(1)} /r_{(1)} , \quad {\bf v}_{(1)} \equiv {\bf \dot{x}}_{(1)}.
\ee
Now, writing $r_{(1)}$, ${\bf n}_{(1)}$ and ${\bf v}_{(1)}$ as first-order expansions in $c^{-2}$, and then inserting (\ref{PNnotationSSS}) and (\ref{activemass-SSS}) into (\ref{PNmotion-testparticle-isotropic-SSS}), one finds easily that the latter decomposes into an equation for the order 0, which is the Newtonian equation ${\bf \ddot{x}} =  -\frac{GM}{r^2} {\bf n}$, and an equation for the order 1 in $c^{-2}$, which is exactly Eq. (\ref{PNcorr-PP-SSS}). This proves that indeed, the 1PN acceleration of a small body is equal, at the point-particle limit, to that of a test particle--- at least in the SSS case.

\section{Conclusion} \label{conclusion} 
The investigated theory starts from a heuristic interpretation \cite{O3} of gravity as the pressure force exerted on the elementary particles by an universal fluid or ``constitutive ether" \cite{Romani}, of which these particles themselves would be just local organizations. This leads naturally to assuming that gravity affects the physical standards of space and time, by an analogy with the effects of a uniform motion that are at the basis of Lorentz-Poincar\'e (special) relativity \cite{O3}. However, the contraction of physical objects in a gravitational field, as it appears in terms of the ``unaffected" Euclidean metric, may either occur in one direction only \cite{A15}, as for the Lorentz contraction, or else \cite{Podlaha,Broekaert03,O3} it may affect all (infinitesimal) directions equally. The first version of the theory was based on a unidirectional contraction, and passed a number of tests \cite{B21,A34}, but it has been discarded by a significant violation of the weak equivalence principle (WEP), which has been found to occur for extended bodies at the point-particle limit \cite{A33}. In the present paper, a new version of the theory, based on a locally {\it isotropic} contraction, has been fully constructed. It has been shown that the new version also explains the gravitational effects on light rays (Subsect. \ref{light-rays}). Being based on the same dynamics as the former version, and being also based on a wave equation for the scalar gravitational field, it should lead, as did the former version \cite{A34}, to a ``quadrupole formula" similar to that used in GR to analyse the data of binary pulsars \cite{TaylorWeisberg}. Moreover, because the metric in the new version is similar to the ``standard PN metric" of GR while the local equations of motion are also similar to those of GR, the celestial mechanics of that theory should improve over Newtonian celestial mechanics.
\footnote{
Of course, the theory is not equivalent to GR, e.g. there is not the Lense-Thirring effect in the usual sense \cite{A19}. However, rotation of a massive body does have dynamical effects in this theory, including effects on that body's own acceleration (as already in the first version \cite{A32}). According to the asymptotic PN scheme that we use, the same type of rotation effect is present also in GR \cite{A36}.
}
These two points will have to be checked in a future work. \\

It has been proved here that the present new version of the theory solves completely the problem with the WEP, that occurred at the first post-Newtonian approximation in the former version (Sect. \ref{pointlimit}). When that (deadly) WEP violation had been found, it had been argued \cite{A33} that the reason for it was the dependence of the PN spatial metric on the spatial derivatives of the Newtonian potential $U$. 
By switching to an isotropic space metric, whose PN form depends on $U$ but not on its derivatives, we indeed suppressed the WEP violation in the present new version.  
%
\bibliography{apssamp}

\end{document}